\documentclass{aa}  

\usepackage{graphicx}
\usepackage{txfonts}
\usepackage[hidelinks]{hyperref}
\usepackage[nameinlink,capitalize]{cleveref}
\usepackage{amsmath}
\usepackage{xcolor}
\usepackage{orcidlink}

\newcommand{\avg}[1]{\left\langle#1\right\rangle}

\renewcommand{\sectionautorefname}{Sect.}
\let\subsectionautorefname\sectionautorefname
\let\subsubsectionautorefname\subsectionautorefname

\begin{document}

\renewcommand{\sectionautorefname}{Sect.}
\let\subsectionautorefname\sectionautorefname
\let\subsubsectionautorefname\subsectionautorefname

\renewcommand*{\figureautorefname}{Fig.}
   \title{COMAP Pathfinder  --  Season 2 results\\III. Implications for cosmic molecular gas content\\at ``Cosmic Half-past Eleven''}
\authorrunning{D.~T.~Chung et al.}
\titlerunning{COMAP Season 2 -- III. Implications for cosmic molecular gas}
   \author{D.~T.~Chung\inst{1}\fnmsep\inst{2}\fnmsep\inst{3}\fnmsep\thanks{\email{dongwooc@cita.utoronto.ca}}\orcidlink{0000-0003-2618-6504}
          \and
          P.~C.~Breysse\inst{4}\fnmsep\inst{5}\orcidlink{0000-0001-8382-5275}
          \and
          K.~A.~Cleary\inst{6}\orcidlink{0000-0002-8214-8265}
          \and
          D.~A.~Dunne\inst{6}\orcidlink{0000-0002-5223-8315}
          \and
          J.~G.~S.~Lunde\inst{7}\orcidlink{0000-0002-7091-8779}
          \and
          H.~Padmanabhan\inst{8}\orcidlink{0000-0002-8800-5740}
          \and
          N.-O.~Stutzer\inst{7}\orcidlink{0000-0001-5301-1377}
          \and
          D.~Tolgay\inst{1}\fnmsep\inst{9}\orcidlink{0000-0002-3155-946X}
          \and 
          J.~R.~Bond\inst{1}\fnmsep\inst{9}\fnmsep\inst{10}\orcidlink{0000-0003-2358-9949}
          \and
          S.~E.~Church\inst{11}
          \and
          H.~K.~Eriksen\inst{7}\orcidlink{0000-0003-2332-5281}
          \and
          T.~Gaier\inst{12}
          \and
          J.~O.~Gundersen\inst{13}\orcidlink{0000-0002-7524-4355}
          \and
          S.~E.~Harper\inst{14}\orcidlink{0000-0001-7911-5553}
          \and
          A.~I.~Harris\inst{15}\orcidlink{0000-0001-6159-9174}
          \and
          R.~Hobbs\inst{16}
          \and
          H.~T.~Ihle\inst{7}\orcidlink{0000-0003-3420-7766}
          \and
          J.~Kim\inst{17}\orcidlink{0000-0002-4274-9373}
          \and
          J.~W.~Lamb\inst{16}\orcidlink{0000-0002-5959-1285}
          \and
          C.~R.~Lawrence\inst{12}\orcidlink{0000-0002-5983-6481}
          \and
          N.~Murray\inst{1}\fnmsep\inst{9}\fnmsep\inst{10}
          \and
          T.~J.~Pearson\inst{6}\orcidlink{0000-0001-5213-6231}
          \and
          L.~Philip\inst{18}\orcidlink{0000-0001-7612-2379}
          \and
          A.~C.~S.~Readhead\inst{6}\orcidlink{0000-0001-9152-961X}
          \and
          T.~J.~Rennie\inst{14}\fnmsep\inst{19}\orcidlink{0000-0002-1667-3897}
          \and
          I.~K.~Wehus\inst{7}\orcidlink{0000-0003-3821-7275}          
          \and
          D.~P.~Woody\inst{16} (COMAP Collaboration)
          }

   \institute{Canadian Institute for Theoretical Astrophysics, University of Toronto, 60 St.~George Street, Toronto, ON M5S 3H8, Canada
   \and Dunlap Institute for Astronomy and Astrophysics, University of Toronto, 50 St.~George Street, Toronto, ON M5S 3H4, Canada
   \and Department of Astronomy, Cornell University, Ithaca, NY 14853, USA
       \and Center for Cosmology and Particle Physics, Department of Physics, New York University, 726 Broadway, New York, NY 10003, USA 
       \and Department of Physics, Southern Methodist University, Dallas, TX 75275, USA
       \and 
       California Institute of Technology, 1200 E. California Blvd., Pasadena, CA 91125, USA
       \and Institute of Theoretical Astrophysics, University of Oslo, P.O. Box 1029 Blindern, N-0315 Oslo, Norway
       \and 
       Departement de Physique Th\'{e}orique, Universite de Gen\`{e}ve, 24 Quai Ernest-Ansermet, CH-1211 Gen\`{e}ve 4, Switzerland
       \and Department of Physics, University of Toronto, 60 St.~George Street, Toronto, ON, M5S 1A7, Canada
       \and David A.~Dunlap Department of Astronomy, University of Toronto, 50 St.~George Street, Toronto, ON, M5S 3H4, Canada
       \and Kavli Institute for Particle Astrophysics and Cosmology \& Physics Department, Stanford University, Stanford, CA 94305, USA
       \and Jet Propulsion Laboratory, California Institute of Technology, 4800 Oak Grove Drive, Pasadena, CA 91109, USA
       \and Department of Physics, University of Miami, 1320 Campo Sano Avenue, Coral Gables, FL 33146, USA
       \and Jodrell Bank Centre for Astrophysics, Department of Physics \& Astronomy, The University of Manchester, Oxford Road, Manchester, M13 9PL, UK
       \and Department of Astronomy, University of Maryland, College Park, MD 20742, USA
       \and Owens Valley Radio Observatory, California Institute of Technology, Big Pine, CA 93513, USA
       \and Department of Physics, Korea Advanced Institute of Science and Technology (KAIST), 291 Daehak-ro, Yuseong-gu, Daejeon 34141, Republic of Korea
       \and Brookhaven National Laboratory, Upton, NY 11973-5000
       \and Department of Physics and Astronomy, University of British Columbia, Vancouver, BC Canada V6T 1Z1, Canada
             }

   \date{Received DD MMM YYYY; accepted DD MMM YYYY}

  \abstract
   {The Carbon monOxide Mapping Array Project (COMAP) Pathfinder survey continues to demonstrate the feasibility of line-intensity mapping using high-redshift carbon monoxide (CO) line emission traced at cosmological scales. The latest COMAP Pathfinder power spectrum analysis is based on observations through the end of Season~2, covering the first three years of Pathfinder operations. We use our latest constraints on the CO(1--0) line-intensity power spectrum at $z\sim3$ to update corresponding constraints on the cosmological clustering of CO line emission and thus the cosmic molecular gas content at a key epoch of galaxy assembly. We first mirror the COMAP Early Science interpretation, considering how Season 2 results translate to limits on the shot noise power of CO fluctuations and the bias of CO emission as a tracer of the underlying dark matter distribution. The COMAP Season~2 results place the most stringent limits on the CO tracer bias to date, at $\avg{Tb}<4.8$\,$\mu$K. These limits narrow the model space significantly compared to previous CO line-intensity mapping results while maintaining consistency with small-volume interferometric surveys of resolved line candidates. The results also express a weak preference for CO emission models used to guide fiducial forecasts from COMAP Early Science, including our data-driven priors. We also consider directly constraining a model of the halo--CO connection, and show qualitative hints of capturing the total contribution of faint CO emitters through the improved sensitivity of COMAP data. With continued observations and matching improvements in analysis, the COMAP Pathfinder remains on track for a detection of cosmological clustering of CO emission.}


   \maketitle
%
\clearpage
\section{Introduction}
\label{sec:intro}
Line-intensity mapping (LIM) surveys map the large-scale structure of the Universe in large cosmological volumes, but not through discrete resolved tracer sources. Rather, LIM surveys achieve this through unresolved emission in specific spectral lines, including lines associated with different phases of the star-forming interstellar medium (ISM) such as carbon monoxide (CO) and the [C\textsc{\,ii}] line from singly ionized carbon (see~\citealt{LIM2019} and \citealt{BernalKovetz22_} for recent reviews). As part of a range of emerging interferometric and single-dish LIM surveys from radio to sub-millimeter wavelengths, the CO Mapping Array Project (COMAP;~\citealt{COMAPESI}) is building a dedicated centimeter-wave LIM program to map the cosmic clustering of emission in the CO(1--0) and CO(2--1) lines from the epochs of galaxy assembly ($z\sim3$, just before so-called ``cosmic noon'') and reionization ($z\sim7$, ``cosmic dawn''). COMAP science will encompass both the astrophysics of the assembly of molecular gas at these key epochs of galaxy evolution, and ultimately the cosmological implications of observed high-redshift large-scale structure traced by CO emission.

The first phase of COMAP is the COMAP Pathfinder, a 26--34\,GHz spectrometer comprising a single-polarization 19-feed array of coherent receivers on a single 10-meter dish at the Owens Valley Radio Observatory~\citep{COMAPESII}. The focus of the Pathfinder survey is on CO(1--0) emission from $z\sim3$, or a lookback time of $\sim11.5$ Gyr. Around this ``cosmic half-past eleven'', we survey galaxies assembling towards the ``cosmic noon'' of peak cosmic star-formation activity~\citep{2015ARA&A..53...51S,2020ARA&A..58..661F}. Following the Early Science analysis of~\cite{COMAPESIII} and~\cite{COMAPESIV} based on the first season of observations (Season 1), the Season 2 data analysis by~\cite{lunde:2024} and~\cite{stutzer:2024} encompasses three years of observations and improved data cleaning methods for almost an order-of-magnitude increase in power spectrum sensitivity.

With such progress continuing to demonstrate the feasibility of CO LIM survey operations and low-level data analysis, we present here the corresponding update on our understanding of CO(1--0) emission at $z\sim3$. We carry out a high-level analysis of the power spectrum constraints of~\cite{stutzer:2024} to answer the following questions:
\begin{itemize}
    \item How much does the increased data volume improve constraints on the clustering and shot noise power of cosmological CO(1--0) emission at $z\sim3$?
    \item Can COMAP Season 2 data better constrain the empirical connection between CO emission and the underlying structures of dark matter?
\end{itemize}

We structure the paper as follows. In~\autoref{sec:methods} we outline our methodology for interpretation, including but no longer limited to methods previously used in~\cite{COMAPESV_}. We discuss the results of our analysis in~\autoref{sec:results}, and implications for understanding CO emission and interpreting past and future CO LIM surveys in~\autoref{sec:discussion}. We end with our primary conclusions and future outlook in~\autoref{sec:conclusions}.

We assume a $\Lambda$CDM cosmology with parameters $\Omega_m = 0.286$, $\Omega_\Lambda = 0.714$, $\Omega_b =0.047$, $H_0=100h$\,km\,s$^{-1}$\,Mpc$^{-1}$ with $h=0.7$, $\sigma_8 =0.82$, and $n_s =0.96$, to maintain consistency with previous COMAP simulations~\citep[starting with][]{Li16}. Distances carry an implicit $h^{-1}$ dependence throughout, which propagates through masses (all based on virial halo masses, proportional to $h^{-1}$) and volume densities ($\propto h^3$). Logarithms are base-10 unless stated otherwise.
\section{Methods}
\label{sec:methods}
The primary target of the COMAP Pathfinder is the power spectrum of spatial-spectral emission after subtraction of continuum emission and systematic effects. Any residual fluctuations should predominantly arise from clustered populations of CO-emitting high-redshift galaxies, meaning that we interpret any constraints on the residual emission as constraints on these CO emitters. In the simplest possible model, the power spectrum as a function of comoving wavenumber $k$ consists of the matter power spectrum $P_m(k)$ scaled by some amplitude $A_\text{clust}$, plus a scale-independent shot noise amplitude $P_\text{shot}$:
\begin{equation}
    P(k) = A_\text{clust}P_m(k) + P_\text{shot}.\label{eq:pk}
\end{equation}
The matter power spectrum describes the distribution of matter density contrast across comoving space, and evolves with redshift as large-scale structure forms and grows. The spatial-spectral fluctuations in CO brightness temperature across cosmological scales trace the clustering of the underlying matter fluctuations with some bias, which informs the clustering amplitude $A_\text{clust}$. In combination with halo models of CO emission that postulate average CO luminosities per halo of collapsed dark matter, constraining $A_\text{clust}$ (or related quantities) and $P_\text{shot}$ allows us to understand not just the global cosmic abundance of CO, but also the relative contribution of different sizes of halos and thus of galaxies. Estimation of the CO line power spectrum $P(k)$ is thus a key target of COMAP low-level analyses.

The goal of this section is to outline methods for the kind of analyses suitable for the current level of sensitivity achieved by the COMAP dataset. First,~\autoref{sec:comapdat} reviews the COMAP Season 2 power spectrum results in relation to previous work. Then,~\autoref{sec:methods_2param} reviews a simple two-parameter analysis as carried out by~\cite{COMAPESV_} for COMAP Early Science, constraining the clustering and shot noise amplitudes and only indirectly using halo models to support physical interpretation. Finally,~\autoref{sec:mcmcmet} outlines a five-parameter analysis to directly constrain a halo model of CO emission, as carried out by~\cite{COMAPESV_} to derive priors for COMAP Early Science but incorporating COMAP data for the first time.

\subsection{Foundational data: COMAP Season 2 power spectrum constraints}
\label{sec:comapdat}
\begin{figure*}
    \centering
    \includegraphics[width=0.986\linewidth,clip=True]{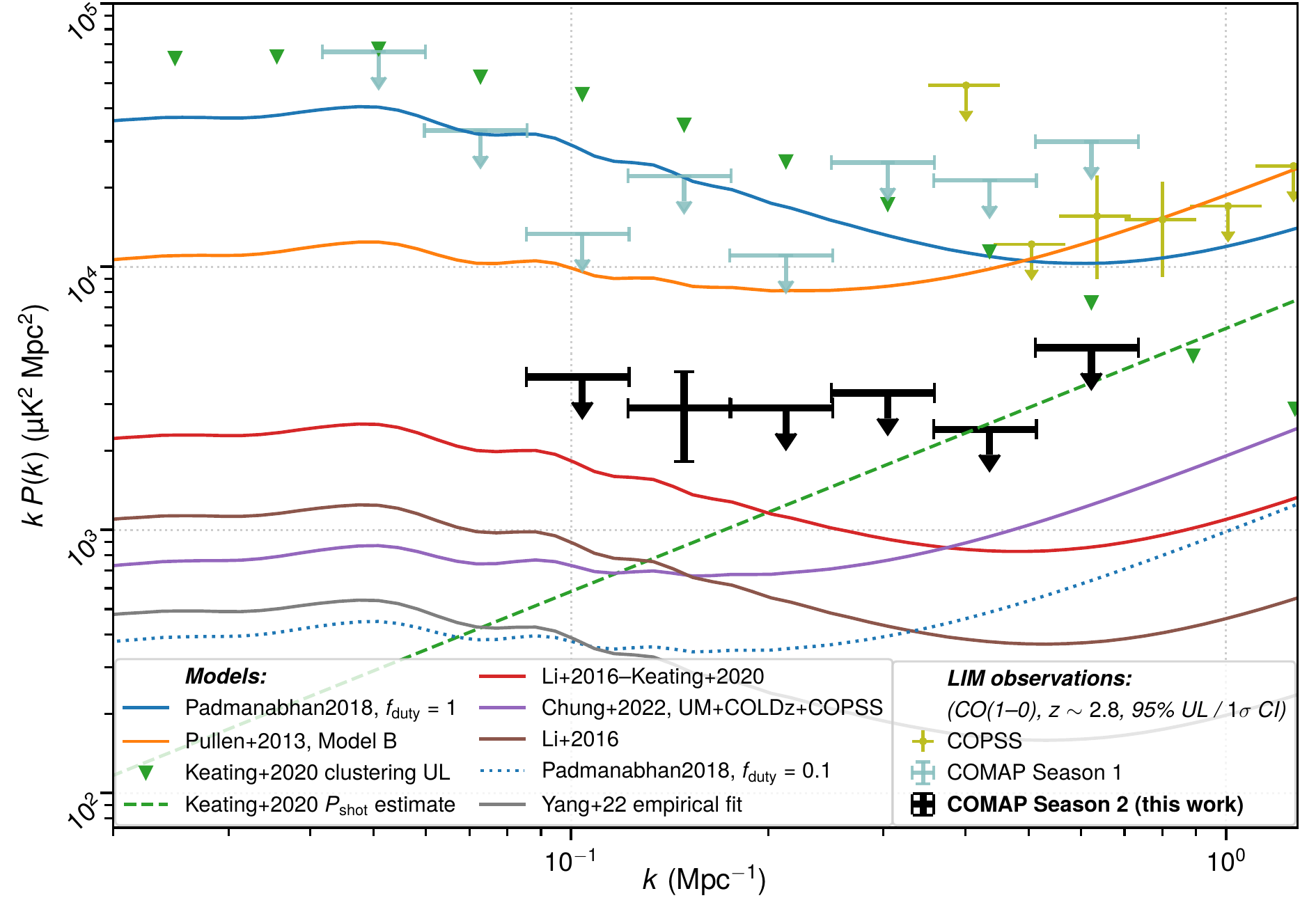}
    \caption{COMAP Season 2 95\% upper limits (given $P(k)>0$) on the $z\sim3$ CO(1--0) power spectrum, with analogous limits from COPSS~\citep{Keating16} and COMAP Season 1~\citep{COMAPESV_}. Some $k$-bins in COPSS and COMAP Season 2 data show marginal excesses, influencing analyses in this work; we thus show $1\sigma$ intervals for these bins unlike in~\cite{stutzer:2024}. We also show predictions based on~\cite{COMAPESV_},~\cite{Padmanabhan18},~\cite{Pullen13},~\cite{Li16}, and~\cite{Yang22}, plus a variation on the~\cite{Li16} model from~\cite{mmIME-ACA}, and the~\cite{mmIME-ACA} re-analysis of COPSS constraining clustering (triangles) and shot-noise amplitudes (dashed line).}
    \label{fig:pk}
\end{figure*}

The present work makes use of the results of \cite{stutzer:2024}, which derived updated power spectrum constraints based on COMAP Pathfinder survey data collected across 17500 hours over three fields of 2--3 deg$^2$ each, between its commissioning in May 2019 and the end of the second observing season in November 2023. We also make use of the prior work of the CO Power Spectrum Survey (COPSS;~\citealt{Keating16}), which performed a pilot CO LIM survey targeting largely the same observing frequencies, but with an interferometric dataset probing smaller scales. The COMAP observations are subject to the effects of instrument and pipeline response, such as filtering, pixelization, and beam smoothing. However, the results as considered in this work correct for these effects by applying the inverse of the estimated power spectrum transfer function per $k$-bin. We expect mode mixing in COMAP data is still at the level of~\cite{COMAPESIV} at most, that is to say less than 20\% for the comoving wavenumber range of $k\gtrsim0.1$\,Mpc$^{-1}$ that we consider.

\autoref{fig:pk} shows these results alongside the range of expectations for the $z\sim3$ CO(1--0) emission power spectrum from empirical modeling in the decade leading up to this dataset~\citep{Pullen13,Li16,Padmanabhan18,mmIME-ACA,COMAPESV_,Yang22}. These models either postulate a connection between dark matter halo properties and CO luminosity via intermediate galaxy properties like star-formation rate (SFR), or directly model the halo--CO connection constrained by observed CO luminosity functions and CO LIM measurements.

Of the models shown in~\autoref{fig:pk}, only the models of~\cite{Padmanabhan18} and~\cite{COMAPESV_} fall into the latter category. \cite{mmIME-ACA} also provide empirical estimates for the clustering and shot noise amplitudes, but this is simply based on decomposing the COPSS measurement of~\cite{Keating16} into clustering and shot noise components, rather than a detailed halo model. In a different context~\cite{mmIME-ACA} do provide a halo model, which we term the~\cite{Li16}--\cite{mmIME-ACA} model, varying the~\cite{Li16} model by using the same halo--SFR connection from~\cite{Behroozi13b,Behroozi13a} but replacing the SFR--CO connection (via infrared luminosity) derived from a compilation of local and high-redshift galaxies \citep{CW13} with one based on a local sample observed by~\cite{Kamenetzky16}.

Even before any detailed analyses, compared to COMAP Season 1 we clearly see an increasing rejection of Model B of~\cite{Pullen13} and of the~\cite{Padmanabhan18} model with CO emission duty cycle $f_\text{duty}=1$. We refer the reader to the Early Science work of~\cite{COMAPESV_} for the implications of excluding these models. As with COMAP Early Science, we exclude these models in the clustering regime, rather than the shot-noise dominated scales surveyed by COPSS. However, the COMAP Season 2 sensitivity is sufficient to exclude these models clearly in \emph{individual} $k$-bins of width $\Delta(\log{k\,\text{Mpc}})=0.155$, rather than having to rely on a co-added measurement across all $k$ as was necessary in Early Science. For reference, we show in~\autoref{sec:unsafe} the original data points behind these upper limits, in a way that more closely resembles Fig.~4 of~\cite{stutzer:2024}.

Note also a weak tension against the previous positive COPSS measurement in overlapping $k$-ranges. The original COPSS analysis of~\cite{Keating16} measured the CO power spectrum at $k=1h$\,Mpc$^{-1}$ to be $P(k)=(3.0\pm1.3)\times10^3h^{-3}\,\mu$K$^2$\,Mpc$^3$, for a best estimate of $P(k=0.7\,$Mpc$^{-1})=8.7\times10^3\,\mu$K$^2$\,Mpc$^3$. This is co-added across the entire $k$-range spanned by COPSS, with the highest sensitivity achieved around $k=0.5h$--$2h$\,Mpc$^{-1}$. By contrast, in a single $k$-bin spanning $k=0.52$--$0.75$\,Mpc$^{-1}$, the present COMAP data places a 95\% upper limit of $7.9\times10^3\,\mu$K$^2$\,Mpc$^3$, lying below the COPSS co-added best estimate. However, the COPSS result is itself only a tentative one at $\approx2.3\sigma$ significance, and so there is no statistically significant discrepancy. COMAP data are also entirely consistent with the estimate of $P_\text{shot}=2.0^{+1.1}_{-1.2}\times10^3h^{-3}\,\mu$K$^2$\,Mpc$^3$ from the later re-analysis of COPSS data by~\cite{mmIME-ACA}, which marginalized over the possible contribution to $P(k)$ from clustering. In fact our power spectrum results show a marginal excess at $k\approx0.15$\,Mpc$^{-1}$ that, while well below the upper limit implied by the direct COPSS re-analysis of~\cite{mmIME-ACA}, does tentatively indicate a preference for models like the~\cite{Li16}--\cite{mmIME-ACA} model. The remainder of this work will establish this preference more quantitatively, and consider other implications of these results.
\subsection{Two-parameter analysis: Constraining CO tracer bias and shot noise}
\label{sec:methods_2param}
The most direct way to analyze the COMAP Season 2 constraints is to decompose the CO power spectrum into clustering and shot noise terms as in~\cref{eq:pk}, with a fixed cosmological model and no assumptions around detailed astrophysical modeling. The COMAP data then constrain the possible range of values for $A_\text{clust}$ and $P_\text{shot}$, which we may then compare to model predictions for these amplitudes for the clustering and shot noise contributions to the power spectrum.

However, physical interpretation requires some amount of guidance from models. Consider a halo model of CO emission where halos of virial mass $M_h$ emit with CO luminosity $L(M_h)$. Suppose that we know the halo mass function $dn/dM_h$ describing the differential number density of halos of mass $M_h$, and the bias $b_h(M_h)$ with which the halo number density contrast traces the continuous dark matter density contrast. Then the cosmological fluctuations in CO(1--0) line temperature trace the underlying dark matter fluctuations with a linear scaling of
\begin{equation}
    \avg{Tb}\propto\int dM_h\,\frac{dn}{dM_h}L(M_h)b_h(M_h).
\end{equation}
This should be understood as a mean line temperature--bias product, with appropriate normalization factors applied to convert luminosity density to brightness temperature. We may also ascribe a dimensionless bias $b$ to CO emission contrast by dividing out the average line temperature or luminosity density:
\begin{equation}
    b=\frac{\int dM_h\,({dn}/{dM_h})\,L(M_h)b_h(M_h)}{\int dM_h\,({dn}/{dM_h})\,L(M_h)}.
\end{equation}
Furthermore, any halo model of $L(M_h)$ will predict the shot noise, proportional to the second bias- and abundance-weighted moment of the $L(M_h)$ function rather than the first moment:
\begin{equation}
    P_\text{shot}\propto\int dM_h\,\frac{dn}{dM_h}L^2(M_h)b_h(M_h).\label{eq:Pshot}
\end{equation}
The quantity $P_\text{shot}$ directly describes the shot noise amplitude, but the same is not true of $\avg{Tb}$ in relation to the clustering amplitude. In real comoving space we would expect $A_\text{clust}=\avg{Tb}^2$, but redshift-space distortions (RSD) enhance the clustering term as large-scale structure coherently attracts galaxies~\citep{Kaiser1987,Hamilton1998}. In the linear regime of small $k$, and given that $\Omega_m(z)\approx1$ at COMAP redshifts,
\begin{equation}A_\text{clust}\approx \avg{Tb}^2\left(1+\frac{2}{3b}+\frac{1}{5b^2}\right).\end{equation}
Based on prior modeling efforts, we consider $b>2$ to be a fairly conservative lower bound on CO tracer bias, as outlined by~\cite{COMAPESV_}. This bound on $b$ in turn allows us to bound $\avg{T}=\avg{Tb}/b$ based on an upper bound on $A_\text{clust}$.

We consider two variants of a two-parameter analysis of the COMAP data, the same carried out in~\cite{COMAPESV_}.
\begin{enumerate}
    \item The first variant is a model-agnostic evaluation of the likelihood of different values of $A_\text{clust}$ and $P_\text{shot}$ given the $P(k)$ data points available from the COPSS results of~\cite{Keating16} and/or from COMAP data through Season 2. We only invoke a conservative limit of $b>2$ to obtain an upper bound on $\avg{T}$ from our constraint on $A_\text{clust}$.
    \item The second variant assumes that given values for $\avg{Tb}^2$ and $P_\text{shot}$, we can expect specific values for $b$ and for an effective line width $v_\text{eff}$ describing the suppression of the high-$k$ CO power spectrum from line broadening~\citep{linewidths_}. Exploration of an empirically constrained model space informs fitting functions for $b$ and $v_\text{eff}$ given only $\avg{Tb}^2$ and $P_\text{shot}$, as provided in Appendix B of~\cite{COMAPESV_}, which then enter into calculation of the redshift-space $P(k)$ accounting for RSD and line broadening. We can directly compare this $P(k)$ to our $P(k)$ data to evaluate the likelihood of different values of $\avg{Tb}^2$ and $P_\text{shot}$. We refer to this variant as the ``$b$- and $v_\text{eff}$-informed'' analysis, versus the first ``$b$- and $v_\text{eff}$-agnostic'' version.
\end{enumerate}
We may then compare likely and unlikely regions of this two-parameter space to model predictions.

\subsection{Five-parameter analysis: Directly constraining the halo--CO connection}
\label{sec:mcmcmet}
Neither variant of our two-parameter analysis truly directly constrains the physical picture of CO emission, only a clustering term and a shot noise term. Given a fixed set of power spectrum measurements, the two-parameter analysis will broadly project likelihood contours favouring either high clustering and low shot noise, or low clustering and high shot noise. Yet physical models should impose a strong prior such that clustering and shot noise co-vary, given that the shot noise also tracks with luminosity density, albeit at a higher order -- cf.~\cref{eq:Pshot}.

Directly modeling and constraining $L(M_h)$ thus has its uses. While dark matter halos are not themselves the direct source of CO emission or indeed any baryonic physics, a halo model of CO emission still serves as a simple way to physically ground interpretation of our CO measurements and introduce priors based on other empirical constraints on the galaxy--halo connection.

\subsubsection{Parameterization and derivation of ``UM+COLDz'' posterior}

To model $L(M_h)$, we use the same parameterization and data-driven procedure as in~\cite{COMAPESV_}. One of the datasets driving this procedure is provided by the CO Luminosity Density at High-$z$ (COLDz) survey~\citep{COLDz,COLDzLF}, which identified line candidates at $z\sim2.4$ through an untargeted interferometric search. In~\cite{COMAPESV_} we also introduced somewhat informative priors based loosely on the work of~\cite{UM}, which devised the~\textsc{UniverseMachine} (UM) framework for an empirical model of the galaxy--halo connection by connecting halo accretion histories to a minimal model of stellar mass growth. We thus once again combine these ``UM'' priors with COLDz data and a basic $L(M_h)$ parameterization, just as in~\cite{COMAPESV_}.

We assume a double power law for the linear average $L(M_h)$. In observer units,
\begin{equation}\frac{L'_\text{CO}(M_h)}{\text{K\,km\,s}^{-1}\text{ pc}^2} = \frac{C}{(M_h/M)^A+(M_h/M)^B}.\label{eq:LM}\end{equation}
For CO(1--0), 
\begin{equation}\frac{L_\text{CO}(M_h)}{L_\odot} = 4.9\times10^{-5}\times\frac{L'_\text{CO}(M_h)}{\text{K\,km\,s}^{-1}\text{ pc}^2}.\end{equation}

We also model stochasticity albeit in a highly simplistic fashion, assuming some level of log-normal scatter $\sigma$ (in units of dex) about the average relation. We inherit this practice from the common use of log-normal distributions to model intrinsic scatter in, e.g., the halo--SFR connection (e.g.:~\citealt{Behroozi13b,Behroozi13a}) and the halo--CO connection as modeled for previous early COMAP forecasts~\citep{Li16}.

The somewhat informative ``UM'' priors for the five free parameters of $L(M_h)$ are as follows:
\begin{align}A&= -1.66\pm2.33,\\B&= 0.04\pm1.26,\\\log{C}&= 10.25\pm5.29,\\\log{(M/M_\odot)}&= 12.41\pm1.77.\end{align}
For log-normal scatter, we assume an initial prior of $\sigma=0.4\pm0.2$ (dex), taking cues from~\cite{Li16} for the central value and slightly broadening the prior.

We then narrow these priors further by matching the luminosity function constraints of the COLDz survey. The matching procedure is similar to that used in~\cite{COMAPESV_}. However, that procedure used a snapshot from the Bolshoi--Planck simulation, used by~\citep{UM} but slightly discrepant against our fiducial cosmology. Here, we use our own peak-patch mocks~\citep{mPP} with virial masses matched to the halo mass function of~\cite{Tinker08}. We extract halos from $z\in(2.35,2.45)$ (or $\chi\in(5720.37, 5844.19)$\,Mpc) from these peak-patch mocks, since we are specifically trying to match a luminosity function constraint at $z\sim2.4$. We thus obtain 161 independent realizations of a halo catalogue from a $1140\times1140\times124$\,Mpc$^3=0.16$ Gpc$^3$ box, comparable to the Bolshoi--Planck snapshot with a box size of $(250/0.678)$\,Mpc (or a volume of $0.05$ Gpc$^3$). A Markov chain Monte Carlo (MCMC) procedure identifies the posterior (narrowed prior) based on a likelihood calculation in addition to the mildly informative priors outlined above. At each step:
\begin{itemize}
    \item We use the sampled $L(M_h)$ parameters to convert a random peak-patch realization of halo masses into CO luminosities.
    \item We then fit a Schechter function to the resulting CO luminosity function of that randomly chosen peak-patch box.
    \item We calculate the log-likelihood by comparing the Schechter function fit against the COLDz posterior for the Schechter function parameters via a kernel density estimator.
\end{itemize}

The MCMC uses 250 walkers for 4242 steps, and we discard the first 1000 steps as a burn-in phase.

The result is an informed distribution, which we call the ``UM+COLDz'' posterior, for the five parameters $\{A,B,C,M,\sigma\}$ of our $L(M_h)$ model.

\subsubsection{Derivation of posteriors incorporating CO LIM data}
To derive posteriors based on CO LIM power spectrum measurements, we rerun the same MCMC procedure used to derive the ``UM+COLDz'' distribution with additional contributions to the likelihood from any discrepancy with the CO LIM results. In other words, we introduce additive log-likelihood terms,
\begin{equation}
    \Delta(\log\mathcal{L})\propto-\sum_k\frac{[P_\text{model}(k)-P_\text{data}(k)]^2}{\sigma^2[P_\text{data}(k)]},
\end{equation}
evaluated against each dataset $P_\text{data}(k)$ with error $\sigma[P_\text{data}(k)]$ for the model $P_\text{model}(k)$ drawn at each MCMC step.\footnote{This approximates the likelihood as Gaussian and independent between $k$-bins, which we consider to be a reasonable approximation at least for COMAP Season 2 data. In obtaining the $P(k)$ result, \cite{stutzer:2024} found that on average, any single $k$-bin correlated with any other $k$-bin at a level of $\lesssim10\%$.}

Using our fiducial cosmology and the halo mass function of~\cite{Tinker08}, we numerically evaluate closed-form expressions describing the CO power spectrum at $z\sim2.8$. We evaluate the above log-likelihood terms against the predicted $P_\text{model}(k)$ without imposing positivity priors, which would be redundant with the always positive predictions of our $P(k)$ model.

We consider three (combinations of) datasets:
\begin{itemize}
    \item The ``UM+COLDz+COPSS'' posterior derives from considering only the addition of COPSS data points as shown in~\cite{Keating16}.
    \item The ``UM+COLDz+COPSS+COMAP S1'' posterior derives from considering both COPSS and COMAP Early Science $P(k)$ constraints from Season 1 data.
    \item The ``UM+COLDz+COPSS+COMAP S2'' posterior derives from considering constraints from both COPSS and the present work on COMAP Pathfinder data through Season 2.
\end{itemize}

While the MCMC procedure itself evaluates posteriors for $\{A,B,C,M,\sigma\}$, we can use the resulting sampling of parameter space to obtain posterior distributions for derived quantities like $L(M_h)$, $\avg{T}$, $b$, and $P_\text{shot}$, and we will look for how (if at all) the ``UM+COLDz+COPSS+COMAP S2'' posterior distinguishes itself from posteriors based on only previous data.
\section{Results}
\label{sec:results}
Having outlined the datasets and methods used in the analyses, we now review the results in relation to previous models and results. We consider outcomes of the two-paramater analysis identifying overall amplitudes for clustering and shot noise power in~\autoref{sec:results_2param}, followed by outcomes of the five-parameter analysis fitting for the $L(M_h)$ relation in~\autoref{sec:mcmcres}.
\subsection{Two-parameter analysis}
\label{sec:results_2param}
\begin{figure}[h!]
    \centering
    \includegraphics[width=0.986\linewidth,clip=True,trim=0 2mm 0 2mm]{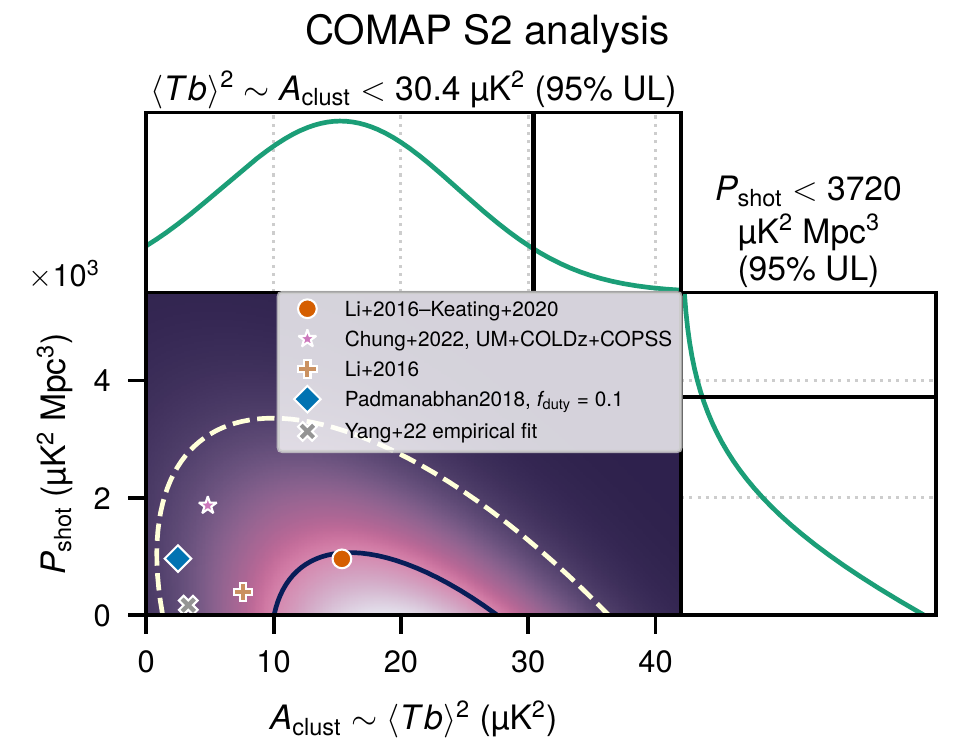}
    \includegraphics[width=0.986\linewidth,clip=True,trim=0 2mm 0 10mm]{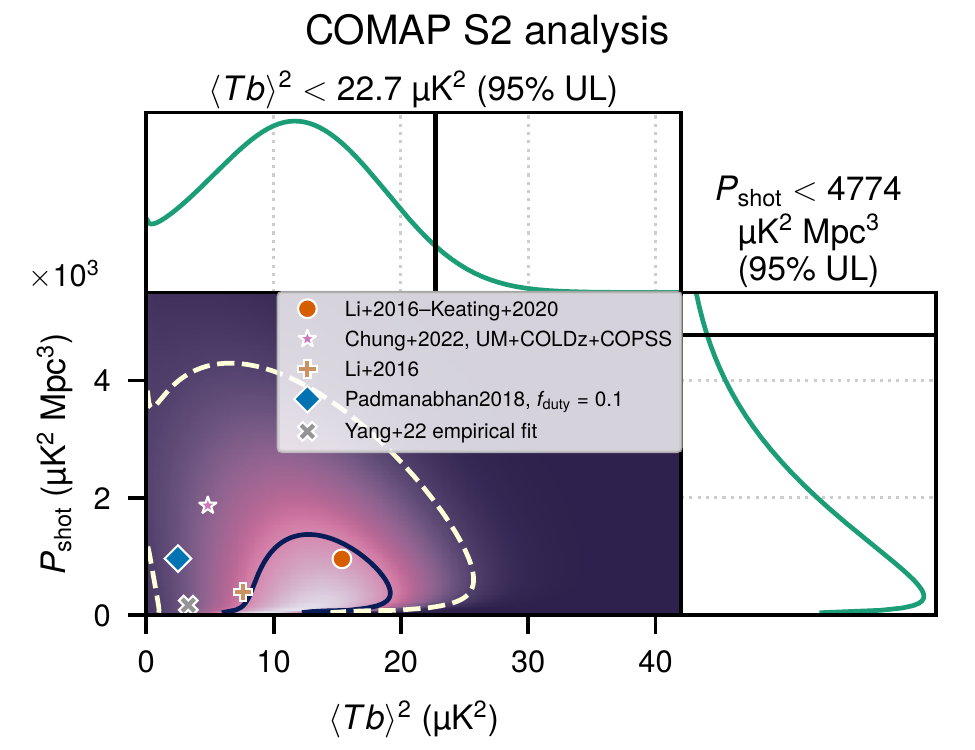}
    \caption{Likelihood contours and marginalized probability distributions for the clustering and shot-noise amplitudes of the CO power spectrum, conditioned on COMAP Season 2 data, in $b$- and $v_\text{eff}$-agnostic (upper) and -informed (lower) analyses. Black solid lines plotted with the 1D marginalized distributions indicate the 95\% upper limits for each parameter. The solid and dashed 2D contours are meant to encompass 39\% and 86\% of the probability mass (delineated at $\Delta\chi^2=\{1,4\}$ relative to the minimum $\chi^2$, corresponding to $1\sigma$ and $2\sigma$ for 2D Gaussians). We show the clustering and shot noise amplitudes for a subset of the models plotted in~\autoref{fig:pk}. Models shown in~\autoref{fig:pk} but not shown here have values of $A_\text{clust}$ or $\avg{Tb}^2$ well beyond the 2$\sigma$ regions shown.}
    \label{fig:2param}
\end{figure}

\begin{figure}[h!]
    \centering
    \includegraphics[width=0.986\linewidth,clip=True,trim=0 2mm 0 2mm]{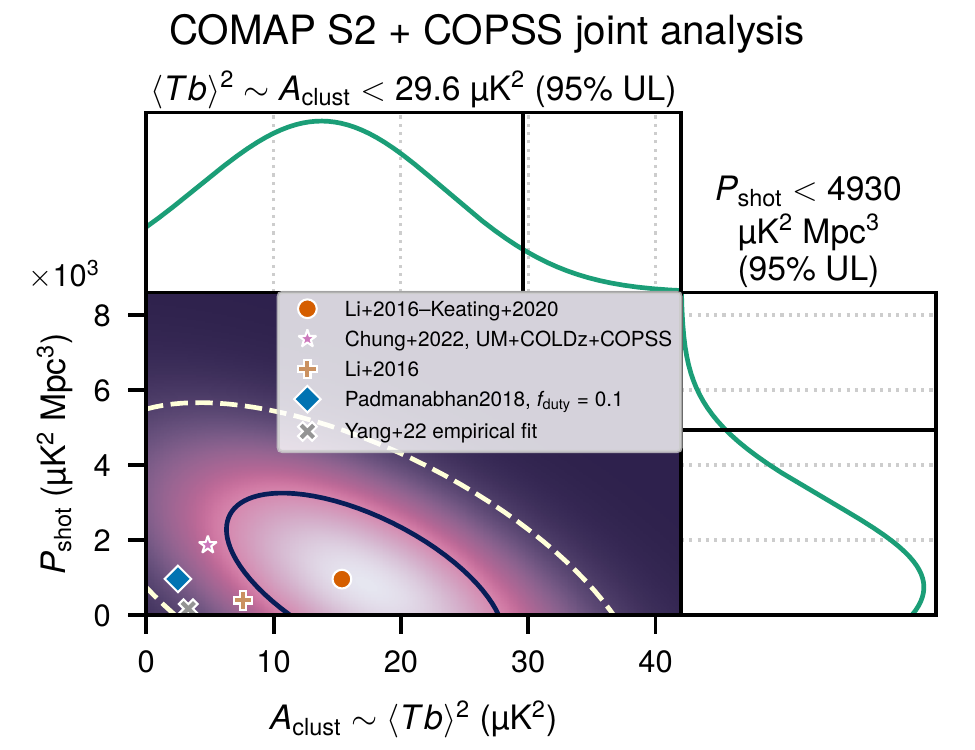}
    \includegraphics[width=0.986\linewidth,clip=True,trim=0 2mm 0 10mm]{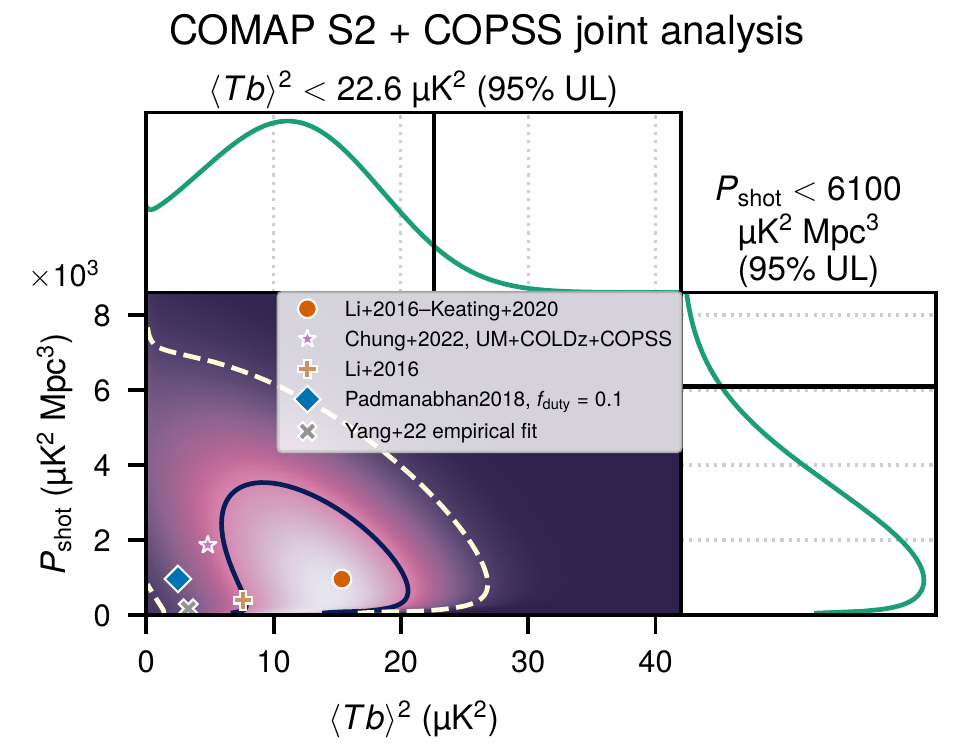}
    \caption{Same as~\autoref{fig:2param}, but with a combination of COMAP Season 2 and COPSS data conditioning the likelihood.}
    \label{fig:2param_copss}
    \end{figure}

\begin{table*}
\centering\begin{tabular}{rcccccccc}
     & \multicolumn{2}{c}{\underline{$b$- and $v_\text{eff}$-agnostic:}}  & \multicolumn{2}{c}{\underline{$b$- and $v_\text{eff}$-informed:}}& \multicolumn{2}{c}{\underline{$b$- and $v_\text{eff}$-agnostic:}}  & \multicolumn{2}{c}{\underline{$b$- and $v_\text{eff}$-informed:}} \\
        &{$A_\text{clust}$}&{$P_\text{shot}/10^3$}&{$\avg{Tb}^2$}&{$P_\text{shot}/10^3$}&{$\avg{T}$}&{$\rho_\text{H2}/10^8$}&{$\avg{T}$}&{$\rho_\text{H2}/10^8$}\\{Data}& {($\mu$K$^2$)}& {($\mu$K$^2$\,Mpc$^3$)}&{($\mu$K$^2$)}&{($\mu$K$^2$\,Mpc$^3$)}&{($\mu$K)}&{($M_\odot\,$Mpc$^{-3}$)}&{($\mu$K)}&{($M_\odot\,$Mpc$^{-3}$)}\\\hline
        COPSS & $<630$ & $5.7^{+4.2}_{-3.6}$ & $<345$ & $12.1^{+7.5}_{-6.4}$ & $<11.$ & $<7.4$ & $<9.3$ & $<6.4$\\
        COMAP S1 & $<66$ & $<19$ & $<49$ & $<24$ & $<3.5$ & $<2.4$ & $<3.5$ & $<2.5$\\
        COMAP S1+COPSS & $<69$ & $6.8^{+3.8}_{-3.5}$ & $<51$ & $11.9^{+6.8}_{-6.1}$ & $<3.5$ & $<2.5$ & $<3.6$ & $<2.5$\\
        {\bf COMAP S2} & ${\bf <31}$ & ${\bf <3.7}$ & ${\bf <23}$ & ${\bf <4.9}$ & ${\bf <2.4}$ & ${\bf <1.6}$ & ${\bf <2.4}$ & ${\bf <1.7}$\\
        COMAP S2+COPSS & $<30$ & $<4.8$ & $<23$ & $<6.1$ & $<2.3$ & $<1.6$ & $<2.4$ & $<1.7$
\end{tabular}
\caption{Results from two-parameter analyses of CO power spectrum measurements for clustering amplitude ($A_\text{clust}$ or $\avg{Tb}^2$) and shot noise power ($P_\text{shot}$), assuming any deviation from zero describes CO(1--0) emission at $z\sim3$. For comparison, we also show results from using only COPSS data or COMAP data through Season 1; we indicate in bold type the results from using COMAP data through Season 2 (without COPSS data). We quote 68\% intervals for $P_\text{shot}$ in the ``COPSS'' and ``COMAP S1+COPSS'' analyses; otherwise we quote 95\% upper limits.}
\label{tab:2param}
\end{table*}

We summarize the results of the two-parameter analysis of the COMAP results in~\autoref{tab:2param}, and show in~\autoref{fig:2param} the probability distributions when considering only COMAP data up to Season 2 (``COMAP S2'' in~\autoref{tab:2param}). We find a factor of 5 improvement in our ability to constrain $P_\text{shot}$ from above with COMAP data alone up to Season 2 compared to COMAP Early Science alone, and a factor of 2 improvement in upper limits for the clustering amplitude. In fact, framing sensitivity to clustering purely in terms of the upper limit achieved downplays our gain. Where the COMAP Early Science analysis effectively gave a maximum \emph{a posteriori} (MAP) estimate of zero for $A_\text{clust}$ and $\avg{Tb}^2$, \autoref{fig:2param} shows that the likelihood distributions peak at positive values of these parameters under COMAP Season 2 constraints.

We also show in~\autoref{fig:2param} model predictions for $A_\text{clust}$ and $P_\text{shot}$, or for $\avg{Tb}^2$ and $P_\text{shot}$. As expected, all models not shown to be excluded by the COMAP Season 2 data at 95\% confidence in~\autoref{fig:pk} are consistent to within $2\sigma$ of the MAP estimate from the COMAP Season 2 likelihood analysis, including the COMAP Early Science fiducial model from~\cite{COMAPESV_}. That said, the most favoured model (within $1\sigma$ of the MAP estimate) is the \cite{Li16}--\cite{mmIME-ACA} model used to explain the results of the mm-wave Intensity Mapping Experiment~\citep[mmIME;][]{mmIME-ACA}. This finding is consistent between the $b$- and $v_\text{eff}$-agnostic and -informed analyses.

The resulting constraints on $\avg{T}$ given $b>2$ are also consistent between these analyses to within a few percent. Going forward we will quote $\avg{Tb}<4.8\,\mu$K or $\avg{T}<2.4\,\mu$K, consistent with both of our ``COMAP S2'' standalone analyses as well as both of the ``COMAP S2+COPSS'' joint analyses as we show in~\autoref{tab:2param}. As in~\cite{COMAPESV_} we can convert any estimate of $\avg{T}$ into an estimate for cosmic molecular gas density:
\begin{equation}
    \rho_\text{H2} = {\alpha_\text{CO}\avg{T}H(z)}/{(1+z)^2}.\label{eq:rhoH2}
\end{equation}
We show the resulting bounds on $\rho_\text{H2}$ in~\autoref{tab:2param} alongside the original bounds on $\avg{T}$, given $\alpha_\text{CO}=3.6\,M_\odot\,($K\,km\,s$^{-1}$\,pc$^2)^{-1}$ and the Hubble parameter $H(z)$ at the central COMAP redshift of $z\sim2.8$. Although some works have advocated for values of $\alpha_\text{CO}$~\citep{Bolatto13,Scoville16} higher by as much as a factor of two, our chosen value follows the one most commonly used by previous CO line search and line-intensity mapping analyses~\citep[e.g.:][]{COLDzLF,Decarli20,PHIBBS2Lenkic,mmIME-ACA}, with this value originally identified in three $z\sim1.5$ galaxies~\citep{Daddi10}.  Our top-line result of $\avg{T}<2.4\,\mu$K corresponds to a bound of $\rho_\text{H2}<1.6\times10^8\,M_\odot$\,Mpc$^{-3}$.

When we use COPSS data in combination with COMAP Season 2 data, the results change to favour higher shot noise values as shown in~\autoref{fig:2param_copss}. The constraints on the clustering amplitude, whether phrased as $A_\text{clust}$ in a $b$-/$v_\text{eff}$-agnostic analysis or $\avg{Tb}^2$ in a $b$-/$v_\text{eff}$-informed analysis, is essentially the same under COMAP Season 2 constraints with or without COPSS data. Note however that in the Early Science analyses of~\cite{COMAPESV_}, the COPSS data dominated the constraint on $P_\text{shot}$ and weakly favoured a positive value, with the $b$/$v_\text{eff}$-agnostic 2D probability distribution between $A_\text{clust}$ and $P_\text{shot}$ resembling a 2D Gaussian distribution just truncated at the $2\sigma$ contour by the $P_\text{shot}=0$ boundary. This is no longer the case, with the corresponding distribution in the upper portion of~\autoref{fig:2param_copss} resembling a 2D Gaussian distribution truncated inside the $1\sigma$ contour.

What greater allowance remains for higher $P_\text{shot}$ values still comes from the way in which the $b$-/$v_\text{eff}$-informed analysis adds an attempted correction for line broadening. Previous work by~\cite{linewidths_} showed that the finite widths of line profiles can attenuate the power spectrum by $\sim10\%$ at scales relevant to COMAP but at a higher $\sim30\%$ level at scales surveyed by interferometric surveys like COPSS. By correcting for this attenuation, the $b$-/$v_\text{eff}$-informed ``COMAP S2+COPSS'' analysis obtains an upper limit of $P_\text{shot}<4.8\times10^3$\,$\mu$K$^2$\,Mpc$^3$, which is 27\% higher than the upper limit from the $b$-/$v_\text{eff}$-agnostic ``COMAP S2+COPSS'' analysis of $P_\text{shot}<6.1\times10^3$\,$\mu$K$^2$\,Mpc$^3$. This difference is within the possible range of attenuation expected for the COPSS $k$-range given our modeling.\footnote{Curiously, however, the upward correction is similar between the COMAP S2 standalone analyses -- in fact, it ends up slightly larger at 32\%. This is not entirely insensible. Even small amounts of attenuation allowable within uncertainties at lower wavenumbers will correspond to a large possible range of corrections for attenuation of the shot noise component, and this lever arm from low $k$ to high $k$ is larger with COMAP data alone than with COPSS data added to the analysis.}

While a combination of low clustering amplitude and high shot noise can certainly explain the current data, the $b$-/$v_\text{eff}$-informed COMAP S2+COPSS analysis shown in~\autoref{fig:2param_copss} assigns significant probability mass within the $2\sigma$ contour to regions of parameter space with high clustering-to-shot noise ratios (particularly at low $\avg{Tb}^2$ values) that do not correspond to any known model. This analysis mode may thus be running into an unphysical parameter space without being grounded in a properly phrased halo model. For instance, by not marginalizing properly over possible values of $v_\text{eff}$, and merely assuming a fixed average for each parameter space point, we potentially incorrectly de-bias against line broadening. We therefore move to consider the five-parameter analysis constraining $L(M_h)$, as opposed to nonspecific clustering and shot noise amplitudes.

\subsection{Five-parameter analysis}
\label{sec:mcmcres}
\begin{figure}
    \centering

    \includegraphics[width=0.986\linewidth,clip=True,trim=0 3mm 0 3mm]{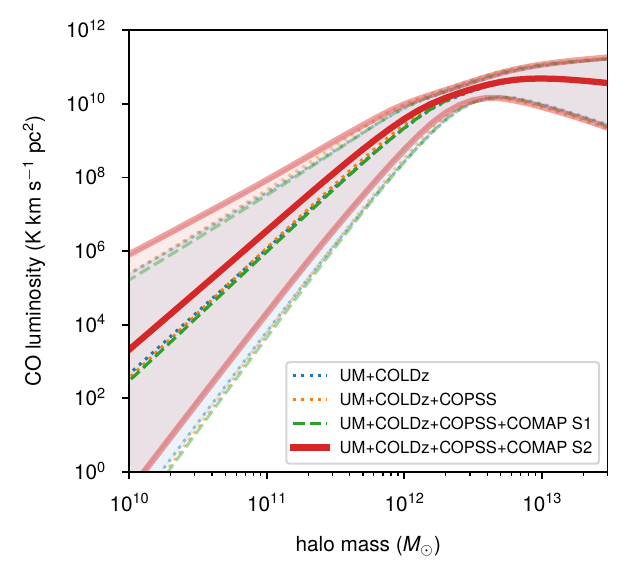}
    \caption{68\% intervals (lighter curves) and median values (darker curves) for $L'_\text{CO}(M_h)$ from the five-parameter MCMC described in the main text.}
    \label{fig:5param_lm}
\end{figure}

\begin{figure}
    \includegraphics[width=0.986\linewidth,clip=True,trim=0 3mm 0 5mm]{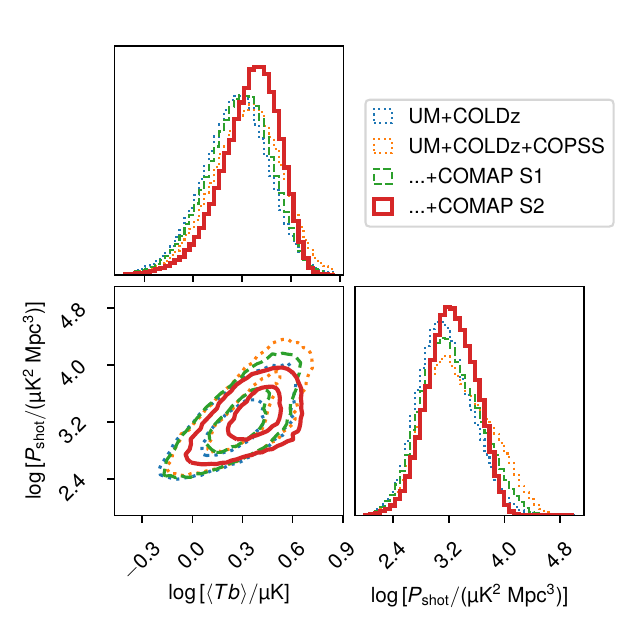}
    \caption{Derived posterior distributions for $\avg{Tb}$ and $P_\text{shot}$ based on the five-parameter MCMC described in the main text. The inner (outer) contours of each distribution show the $1\sigma$ ($2\sigma$) confidence regions.}
    \label{fig:5param_tbshot}
\end{figure}

\autoref{fig:5param_lm} and~\autoref{fig:5param_tbshot} summarize the results of our five-parameter analysis in terms of derived quantities; we also show the posterior distributions in the original parameter space in~\autoref{sec:unexciting}. First, comparing the ``UM+COLDz'' distribution with ``UM+COLDz+COPSS+COMAP S1'', we do not see tangible differences in the derived quantities. The COPSS data on their own push the parameter space towards slightly higher $\sigma$, lower $A$ (so a steeper faint-end slope for the $L(M_h)$ relation), and overall a brighter signal as shown in the posteriors for $\avg{Tb}$ and $P_\text{shot}$. However, the COMAP Season 1 non-detection essentially reverses many of these changes, even suggesting a slightly dimmer faint end of the halo mass--CO luminosity relation.

The COMAP Season 2 results push expectations for the signal back up, albeit only marginally. By pushing the double power law pivot mass $M$ to lower values and pushing $A$ (the opposite of the faint-end slope) to higher values, the COMAP Season 2 analysis suggests a brighter population of low-mass halos than our previous data would allow, as shown in~\autoref{fig:5param_lm}. This is also apparent in~\autoref{fig:5param_tbshot}, where the ``UM+COLDz+COPSS+COMAP S2'' posteriors for the derived quantities $\avg{Tb}$, and $P_\text{shot}$ show a systematic shift towards higher $\avg{Tb}$ -- for the first time markedly pushing away from the lower limit implied by the UM+COLDz distribution -- in addition to a less extended right tail for $P_\text{shot}$. These shifts in the posterior distribution suggest that the COMAP Pathfinder survey is approaching the point of making statements about the faint end of the CO luminosity function by accessing its contribution to the clustering of CO emissivity on cosmological scales, something no other survey on the horizon will do.

\begin{figure}
    \centering
    \includegraphics[width=0.96\linewidth]{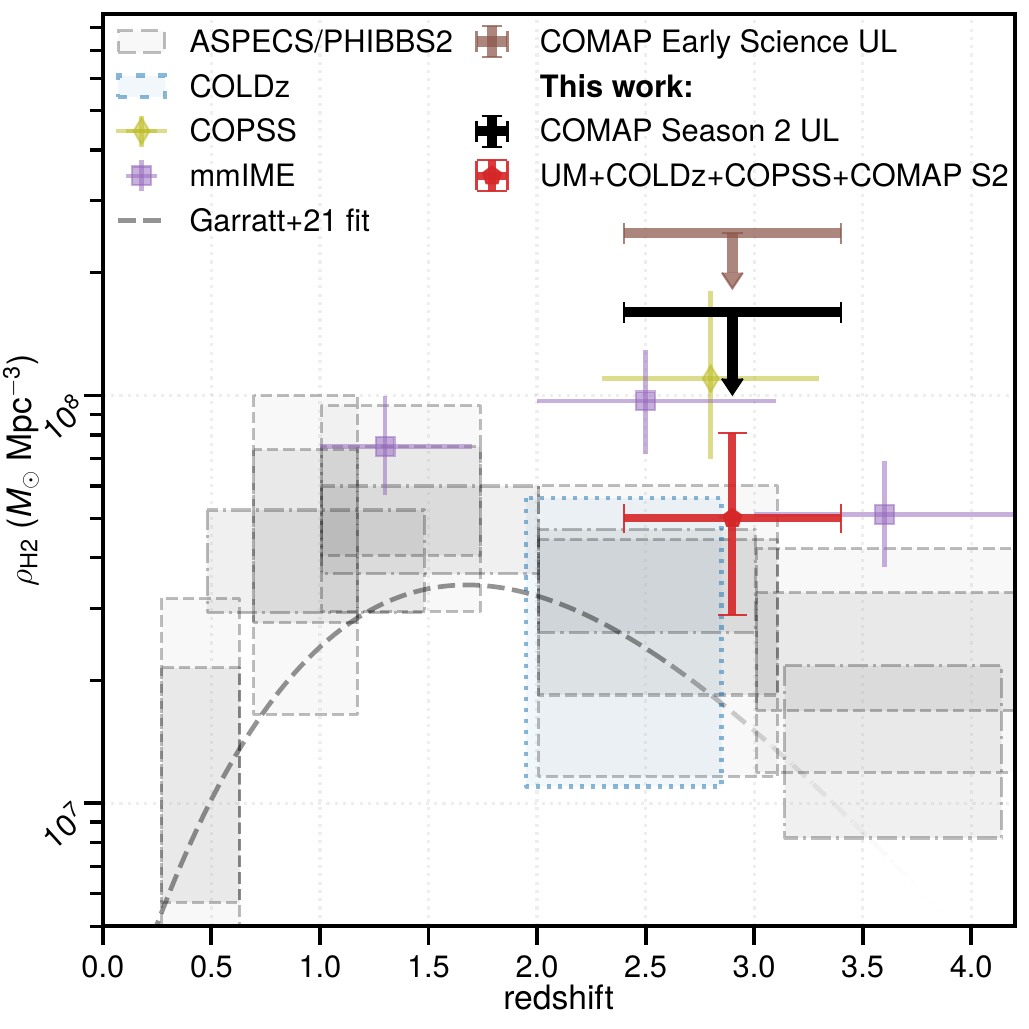}
    \caption{Constraints on the cosmic molecular gas mass density $\rho_\text{H2}$ based on CO abundance measurements across redshifts 0--4. We show COMAP Season 1 and Season 2 constraints on $\avg{T}$ converted based on~\cref{eq:rhoH2}, including both the two-parameter analysis upper limit and the five-parameter UM+COLDz+COPSS+COMAP S2 result. For comparison, we also show past results from untargeted interferometric CO line searches (boxes) -- ASPECS~\citep{Decarli20}, PHIBBS2~\citep{PHIBBS2Lenkic}, and COLDz~\citep{COLDzLF} -- as well as interferometric CO LIM surveys (uncapped error bars) -- COPSS~\citep{Keating16} and mmIME~\citep{mmIME-ACA}. In addition, we show a best-fit model describing results from stacked 850\,$\mu$m luminosities of galaxies at redshifts 0--2.5~\citep{Garratt21}.  All results use $\alpha_\text{CO}=3.6\,M_\odot\,($K\,km\,s$^{-1}$\,pc$^{-2})^{-1}$ (cf.~\citealt{Daddi10}) except COPSS, which uses a conversion of $\alpha_\text{CO}=4.3\,M_\odot\,($K\,km\,s$^{-1}$\,pc$^{-2})^{-1}$ (cf.~\citealt{Bolatto13}), and~\cite{Garratt21}, who use $\alpha_\text{CO}=6.5\,M_\odot\,($K\,km\,s$^{-1}$\,pc$^{-2})^{-1}$ as promoted by~\cite{Scoville16}.}
    \label{fig:rhoH2}
\end{figure}

Finally, we note that the ``UM+COLDz+COPSS+COMAP S2'' estimate for $\avg{T}$ -- which incorporates prior information and should not be considered a COMAP ``detection'' of any kind -- is $0.72^{+0.45}_{-0.30}$\,$\mu$K. This corresponds to a cosmic molecular gas density of $\rho_\text{H2}=5.0^{+3.1}_{-2.1}\times10^7\,M_\odot$\,Mpc$^{-3}$, which we discuss further in the next section.
\section{Discussion}
\label{sec:discussion}

Phrased in terms of constraints on $\rho_\text{H2}$, the COMAP Season 2 results show the progress that COMAP -- and thus single-dish CO LIM as a technique -- has made in growing into an independent probe of cosmological CO emission and thus of cosmic molecular gas content. We illustrate this graphically in~\autoref{fig:rhoH2}, showing the present work's COMAP results in the context of previous work. The results from prior literature are mostly the same as those~\cite{COMAPESV_} collated for their Fig.~9.

\begin{itemize}
    \item Deep surveys have leveraged community interferometers to observe pencil beam volumes and identify CO line emission candidates from the integrated data cubes in a serendipitous fashion. Of surveys used for such untargeted line searches, only the COLDz survey previously discussed in~\autoref{sec:mcmcmet} directly observes high-redshift CO(1--0) line emission.
    \item Two other deep interferometric surveys -- the ALMA SPECtroscopic Survey in the Hubble Ultra Deep Field (ASPECS;~\citealt{Decarli20}) and the Plateau de Bure High-z Blue Sequence Survey 2 (PHIBBS2;~\citealt{PHIBBS2Lenkic}) -- include 3\,mm observations sensitive to a range of CO lines including CO(3--2) at $z\approx2$--3. These constraints on CO luminosity density, and thus $\rho_\text{H2}$, are subject to an additional conversion to CO(1--0) luminosity from higher-$J$ CO lines.
    \item Community interferometers have also hosted key pilot small-scale CO LIM surveys, namely the previously mentioned COPSS and mmIME.
\end{itemize}
As an additional reference point, we also overplot the best-fitting model from \cite{Garratt21} to stacked 850\,$\mu$m luminosities of near-infrared selected galaxies at redshifts 0--2.5. That work took advantage of a tight empirical correlation identified by~\cite{Scoville16} between the 850\,$\mu$m luminosity and CO(1--0) luminosity of both low-redshift galaxies and $z\sim2$ sub-millimeter galaxies. This stands in contrast to the other results assembled, which directly survey CO lines in some fashion, although not always specifically CO(1--0).

While COMAP Season 2 data are in weak disagreement with the COPSS results, this does not translate into a disagreement in the space of $\rho_\text{H2}$. This is due to the way in which \cite{Keating16} derived $\rho_\text{H2}$ from the COPSS results. The derivation involved a number of stringent model assumptions including a linear relation between halo mass and CO luminosity, a linear relation between halo mass and molecular gas mass fraction, and the introduction of a prior on the log-normal scatter $\sigma$ that suppressed the preferred amount of CO luminosity per halo mass versus what an unconstrained analysis would have found. Such assumptions motivate the analyses carried out in the present work, analysing multiple datasets through common modeling frameworks with shared assumptions. Compare, for instance, our own COPSS re-analysis which found an upper limit of $\rho_\text{H2}<6.4$--$7.4\times10^8\,M_\odot$\,Mpc$^{-3}$ in~\autoref{sec:results_2param}, versus our own COMAP S2 upper limit of $\rho_\text{H2}<1.6\times10^8\,M_\odot$\,Mpc$^{-3}$.

\begin{figure}
    \includegraphics[width=0.986\linewidth]{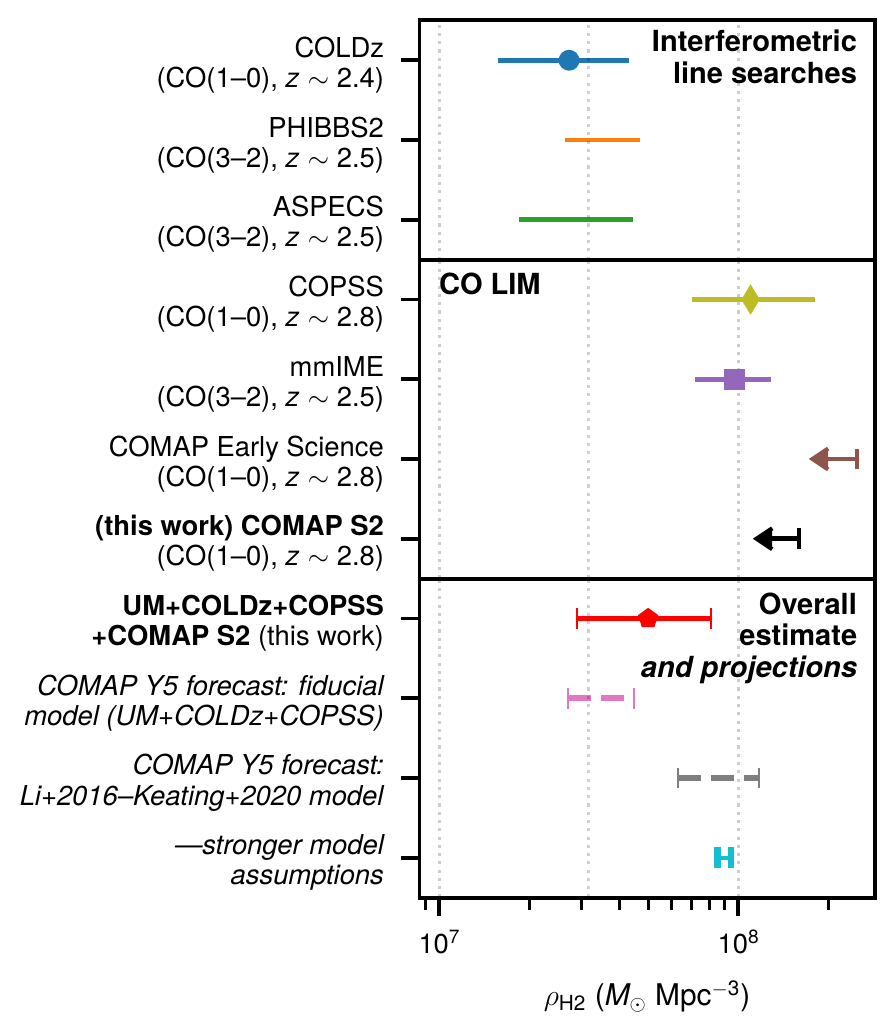}
    \caption{Constraints on $\rho_\text{H2}$ just before ``cosmic noon'', showing 95\% upper limits and confidence intervals standardized to 68\% when possible. In addition to constraints shown before in~\autoref{fig:rhoH2} from line searches~\citep{COLDzLF,PHIBBS2Lenkic,Decarli20}, previous CO LIM analyses~\citep{Keating16,mmIME-ACA,COMAPESV_}, and the present work, we also show COMAP Pathfinder five-year forecasts from~\cite{COMAPESV_}.}
    \label{fig:rhoH2_future}
\end{figure}

As mentioned at the end of~\autoref{sec:mcmcres}, our best estimate for $\rho_\text{H2}$ when combining COMAP Season 2 with external prior information is $\rho_\text{H2}=5.0^{+3.1}_{-2.1}\times10^7\,M_\odot$\,Mpc$^{-3}$. We show in~\autoref{fig:rhoH2} that this ``UM+COLDz+COPSS+COMAP S2'' estimate lies squarely between constraints from CO line searches, which cluster lower, and constraints from interferometric CO LIM surveys, which cluster higher. These two different families of experiments informed two different sets of forecasts of COMAP Pathfinder five-year results in~\cite{COMAPESV_}, one using the fiducial data-driven ``UM+COLDz+COPSS'' model\footnote{Although this model originated from a data-driven prior that also used COPSS, the COLDz data clearly dominated the information content reflected in the prior. We see this again in the present work from the minimal difference between the ``UM+COLDz'' and ``UM+COLDz+COPSS'' posteriors in~\autoref{sec:mcmcres}.} and the other using the \cite{Li16}--\cite{mmIME-ACA} model originally formulated to explain mmIME results. Our best estimate thus also lies between these two models from COMAP Early Science and the forecasts that used them, which we show alongside previous and current results in~\autoref{fig:rhoH2_future}.

As COMAP continues to move forward as a single-dish experiment, its large-scale imaging will complement interferometric surveys in important ways. This includes not only COPSS and mmIME but also resolved line candidate searches like ASPECS, since COMAP relies on statistical large-scale fluctuations rather than individual sources. For example, ASPECS-Pilot detected 14 high-redshift [C\,\textsc{ii}] line candidates~\citep{ASPECSPilot_CII} only to show in the subsequent ASPECS Large Program observations that every single one was spurious~\citep{ASPECSLP_CII}. The importance of having independent single-dish LIM experiments like COMAP in the conversation will only increase as COMAP accrues further data.

The other important focus of COMAP that is salient to the wider landscape of CO abundance measurements is its focus on low-$J$ CO lines, specifically CO(1--0) at $z\sim3$ in the case of the Pathfinder survey. For example, comparing the results of~\cite{Garratt21} against those of ASPECS or PHIBBS2 would require accounting for not only uncertainties in quantities like  $\alpha_\text{CO}$, but also the respective conversion from the original measurement into CO(1--0) luminosity density -- the~\cite{Scoville16} 850\,$\mu$m--CO(1--0) conversion in the case of~\cite{Garratt21}, and the conversion to CO(1--0) from higher-$J$ CO lines observed by ASPECS and PHIBBS2 (though for ASPECS see~\citealt{VLASPECS}). Future COMAP constraints will entirely bypass this last uncertainty by directly constraining the CO(1--0) luminosity density -- and across all faint and bright galaxies in the survey volume, not constrained to any specific galaxy selection.

For now our best estimates remain consistent with all experiments, but our sensitivity to the clustering of CO has clearly improved to the point of providing upward revisions to expectations for the average CO luminosities of low-mass halos. While the current sensitivities of COMAP data to the tracer bias and average line temperature are at best marginal against our informative priors, they will continue to grow as we accrue more data. As~\cite{lunde:2024} note and as we have already noted in the \hyperref[sec:intro]{Introduction}, the COMAP Pathfinder achieved nearly an order-of-magnitude increase in power spectrum sensitivity per $k$-bin despite only a $3.4\times$ increase in raw data volume. With continued improvements in data cleaning and analysis, we remain optimistic that the amount of usable data will increase nonlinearly with further observing seasons and allow the COMAP Pathfinder to meet its targets for five-year sensitivities. As it does so, the resulting constraints will readily lend itself to very straightforward joint analyses with other measurements of cosmic CO(1--0) emissivity or molecular gas content in the vein of other LIM surveys, line scan surveys, or even analyses like that of~\cite{Garratt21}, enhancing our understanding of how the rise and fall of cosmic star-forming gas relates to its depletion through the rise and fall of cosmic star-formation activity.
\section{Conclusions}
\label{sec:conclusions}

With the above results and discussion, we now have firm answers to the questions posed in the \hyperref[sec:intro]{Introduction} to this work:
\begin{itemize}
    \item \emph{How much does the increased data volume improve constraints on the clustering and shot noise power of cosmological CO(1--0) emission at $z\sim3$?} The COMAP Season 2 dataset represents a five-fold improvement in upper bounds on CO shot noise power and a halving of the upper bound on the CO clustering amplitude over COMAP Early Science. This increased sensitivity introduces tension against the previous COPSS result, which will evolve with future analyses.
    \item \emph{Can COMAP Season 2 results better constrain the empirical connection between CO emission and the underlying structures of dark matter?} While COMAP Season 2 data only provide marginal improvements in constraining this connection, we see hints of the COMAP Pathfinder's basic capability in capturing the clustering of low-mass CO emitters in ways that other experiments cannot.
\end{itemize}

The present work has taken an extremely conservative approach to high-level analysis, with generic models for either the power spectrum or the halo--CO connection. By making even stronger model assumptions we can make statements about semi-analytic models of galaxy formation and the connection between star-formation activity and molecular gas content (cf.~\citealt{Breysse22b}). We leave this to a future collaboration work currently in preparation.

The outlook for the COMAP Pathfinder remains strongly positive as it continues past three years of data acquisition. The improvements demonstrated in Season 2, not only in observing efficiency but also in data cleaning and processing as demonstrated by the papers that this work accompanies~\citep{lunde:2024,stutzer:2024}, will continue to grow with further Pathfinder operations. The collaboration thus continues to be on track for the outcome forecast by~\cite{COMAPESV_}: a high-significance detection of cosmological CO clustering sometime in the next few years.

\begin{acknowledgements}
      DTC was supported by a CITA/Dunlap Institute postdoctoral fellowship for much of this work. The Dunlap Institute is funded through an endowment established by the David Dunlap family and the University of Toronto. The University of Toronto operates on the traditional land of the Huron-Wendat, the Seneca, and most recently, the Mississaugas of the Credit River; DTC is grateful to have the opportunity to work on this land. Research in Canada is supported by NSERC and CIFAR.

This material is based upon work supported by the National Science Foundation under Grant Nos.\ 1517108, 1517288, 1517598, 1518282, 1910999, and 2206834, as well as by the Keck Institute for Space Studies under ``The First Billion Years: A Technical Development Program for Spectral Line Observations''.

Parts of the work were carried out at the Jet Propulsion Laboratory, California Institute of Technology, under a contract with the National Aeronautics and Space Administration.

 PCB acknowledges support from the James Arthur Postdoctoral Fellowship during the writing of this work.
     
      We acknowledge and thank the support from the Research Council of Norway through grants 251328 and 274990, and from the European Research Council (ERC) under the Horizon 2020 Research and Innovation Program (grant agreements 772253 bits2cosmology and 819478 Cosmoglobe).

HP acknowledges support from the Swiss National Science Foundation
via Ambizione Grant PZ00P2\_179934.

      This work was supported in part by the National Research Foundation of Korea (NRF) grant funded by the Korean government (MSIT, No.~RS-2024-00340759).

This study uses the Scientific color map \texttt{acton}~\citep{Crameri2018} to prevent visual distortion of the data and exclusion of readers with color-vision deficiencies~\citep{Crameri2020}.

      As in~\cite{COMAPESV_} we thank Riccardo Pavesi for access to the COLDz ABC posterior sample used in this work. Many thanks also to George Stein for running and making available the original peak-patch simulations for use with the COMAP Pathfinder survey. This research made use of NASA's Astrophysics Data System Bibliographic Services. Some of the methods for this work were originally tested on the Sherlock cluster; DTC would like to thank Stanford University and the Stanford Research Computing Center for providing computational resources and support that contributed to these research results.  

      This work was first presented at the Line Intensity Mapping 2024 conference held in Urbana, Illinois; we thank Joaquin Vieira and the other organizers for their hospitality and the participants for useful discussions.
\end{acknowledgements}

\bibliographystyle{aa}
\bibliography{biblio,bibliold,biblionew,references_s2}

\begin{appendix}
\renewcommand{\sectionautorefname}{Sect.}
\let\subsectionautorefname\sectionautorefname
\let\subsubsectionautorefname\subsectionautorefname
\section{Alternative presentation of observed power spectra}
\label{sec:unsafe}
\begin{figure*}
    \centering\setlength{\lineskip}{0pt}
    \includegraphics[width=0.986\linewidth]{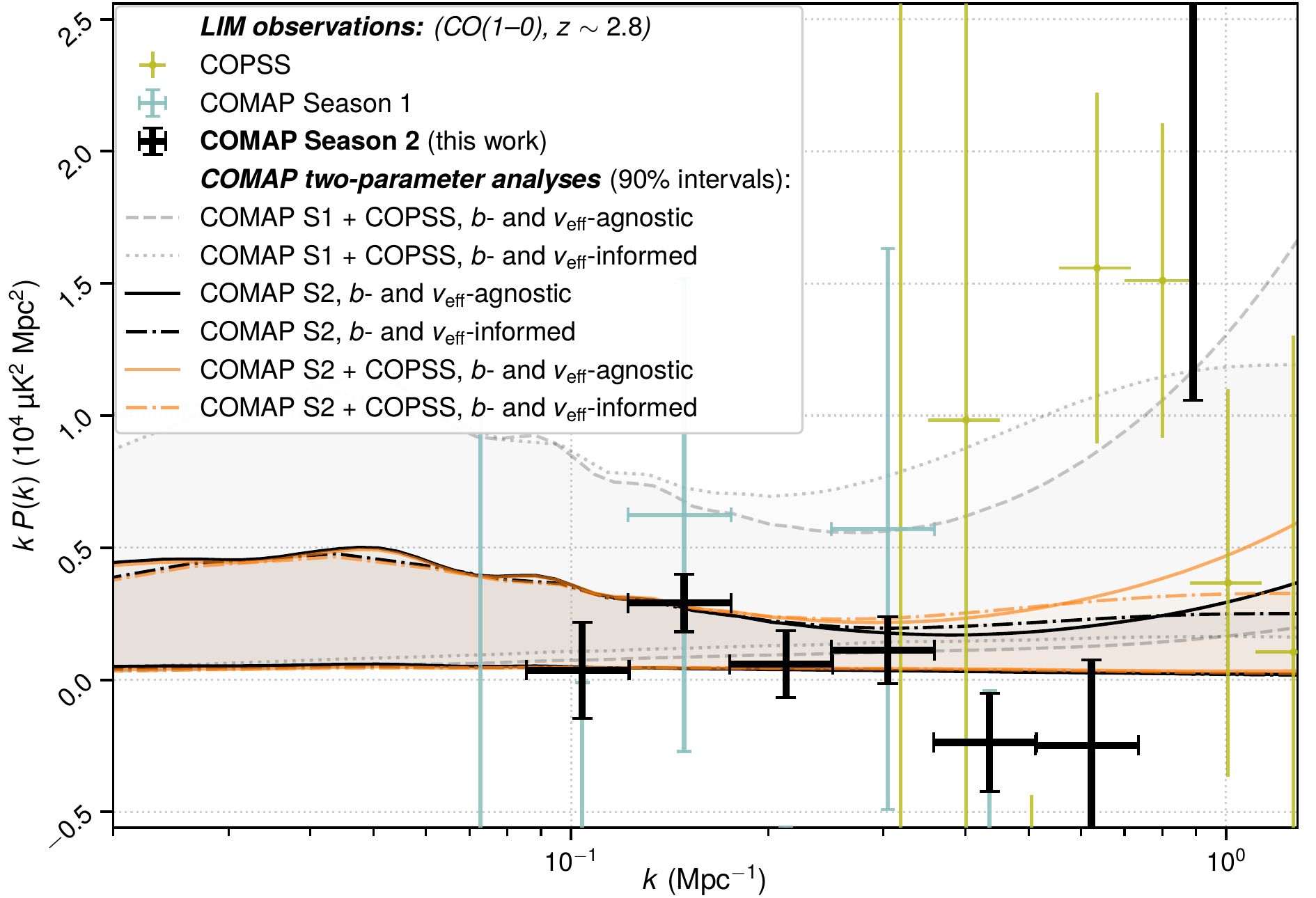}
    \caption{Same as~\autoref{fig:pk}, but with all LIM $P(k)$ results shown with $1\sigma$ uncertainties per $k$-bin and on a linear scale. Furthermore, instead of the specific models shown in~\autoref{fig:pk}, we show the typical range of allowable power spectrum values based on the two-parameter analyses of~\autoref{sec:results}, with the $b$-/$v_\text{eff}$-informed variations showing attenuation for line broadening left uncorrected. These allowable ranges shown should not be taken to represent a detection as they assume non-negative $P(k)$ values by definition.}
    \label{fig:pk_unsafe}
\end{figure*}

\autoref{fig:pk_unsafe} provides an alternate presentation of the LIM observations shown in~\autoref{fig:pk} as part of~\autoref{sec:comapdat}. Namely, we represent all $P(k)$ datasets not as upper limits given positivity priors, and instead as data points on a linear scale. In this case we take the feed or feed-group pseudo cross-power spectrum data $\tilde{C}(k)$ derived in~\cite{COMAPESIV} and~\cite{stutzer:2024} as the best COMAP Season 1 and Season 2 estimates for the astrophysical $P(k)$. We also show credible ranges of $P(k)$ values from the two-parameter analyses of~\autoref{sec:results} rather than specific models as in~\autoref{fig:pk}.

While the Early Science work of~\cite{COMAPESV_} represented COMAP Season 1 and COPSS data as co-added results across their respective $k$-ranges, we will not adopt such representation in future work. Inter-bin correlations, interacting with imposition of a positivity prior on the co-added result versus on the individual $P(k)$ points, could result in differences between co-added and per-bin results impossible to make sense of. For COMAP Season 2 data, we have explicitly verified that inter-bin correlations are $\lesssim10\%$ for our chosen $k$-binning~\citep{stutzer:2024}. Even so, the choice of how to represent a co-added result is fraught with many choices with respect to the averaging scheme, the central $k$-value, and can lead to misleading visual comparisons when plotting model power spectra alongside co-added results without the same weighting used to average the observed power spectra.
\section{Full five-parameter MCMC posterior distributions}
\label{sec:unexciting}
\begin{figure*}
    \includegraphics[width=0.986\linewidth]{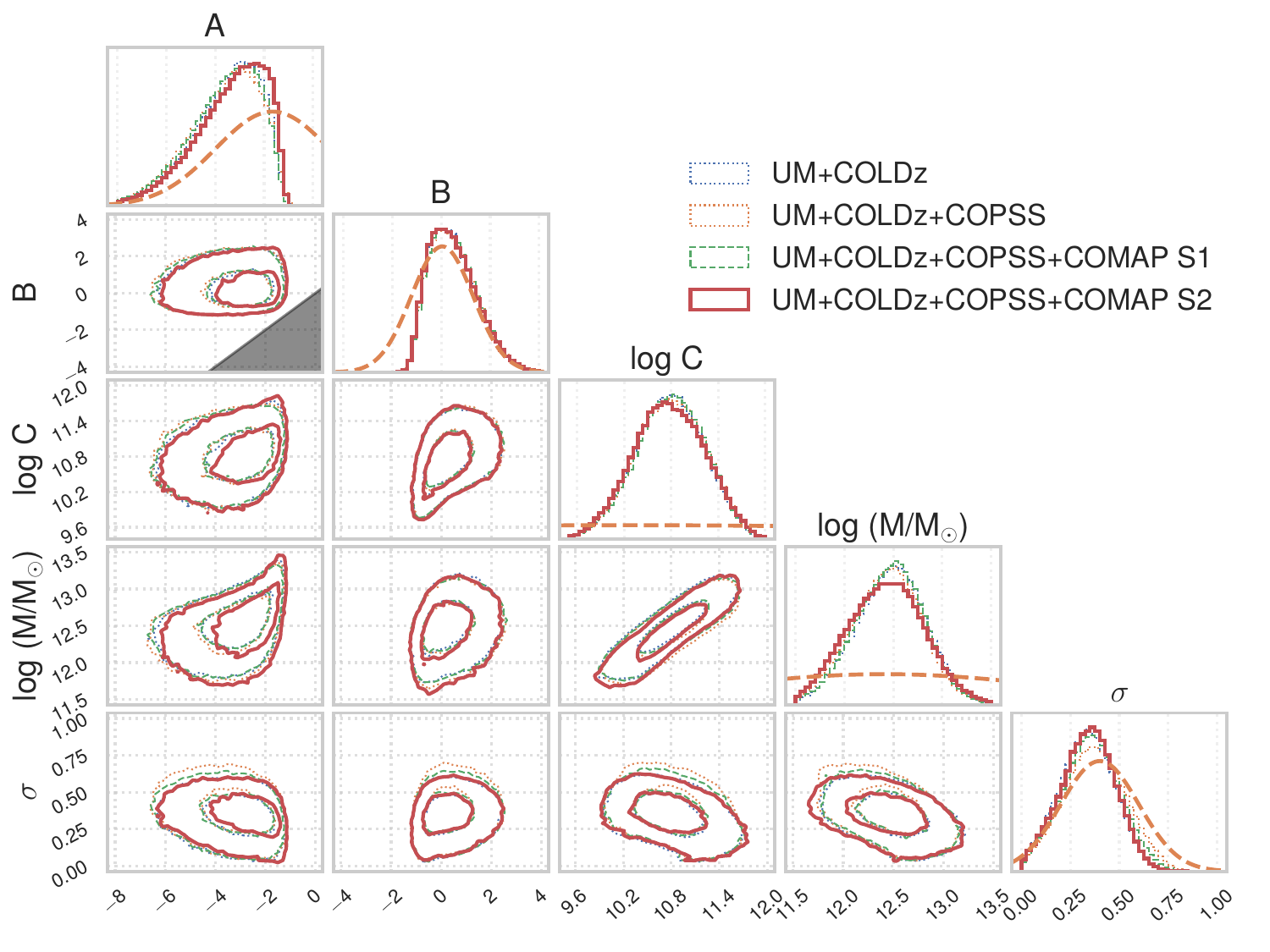}
    \caption{MCMC posterior distributions for the five parameters of our $L'_\text{CO}(M_h)$ model, obtained from the analysis described in~\autoref{sec:mcmcmet} of the main text. The inner (outer) contours of each 2D distribution show the 39\% (86\%) or roughly $1\sigma$ ($2\sigma$) confidence regions. The grey triangle in the 2D probability distribution between $A$ and $B$ shows a never-accessed region where $A>B$; the MCMC treats the two parameters as an interchangeable pair, with the smaller (larger) of the two always subjected to the prior for $A$ ($B$). Dashed lines represent the loosely informative ``UM'' priors discussed in~\autoref{sec:mcmcmet}.}
    \label{fig:5param_mcmcfull}
\end{figure*}

In~\autoref{fig:5param_mcmcfull} we show the full five-parameter posterior distributions from the analysis of~\autoref{sec:mcmcmet}, from which we obtain distributions for the derived quantities shown in~\autoref{sec:mcmcres}. Compared to~\autoref{fig:5param_tbshot}, the changes due to COMAP Season 2 data are more subtle in this higher-dimensional space but nonetheless discernible.

\end{appendix}
\end{document}